\documentclass{article}
\usepackage[margin=1.25in]{geometry}
\usepackage{graphicx}
\usepackage{amsmath}
\usepackage{amssymb}
\usepackage{mathtools}
\usepackage{parskip}
\usepackage[hidelinks]{hyperref}
\usepackage{graphicx}
\usepackage[bf]{caption}
\usepackage{subcaption}

\usepackage{color}
\usepackage{amssymb}
\usepackage{natbib}
\newtheorem{remark}{Remark}[section]
\emergencystretch=\maxdimen
\begin{document}
%%%%%%%%%%%%%%%%%%%%%%%%%
\title{
	Modelling of cyclist's power to overcome dissipative forces on a velodrome
}
%%%%%%%%%%%%%%%%%%%%%%%%%
\author{
	Len Bos%
	\footnote{
		Universit\`a di Verona, Italy, \texttt{leonardpeter.bos@univr.it}
	}\,,
	Michael A. Slawinski%
	\footnote{
		Memorial University of Newfoundland, Canada, \texttt{mslawins@mac.com}
	}\,,
	Rapha\"el A. Slawinski%
	\footnote{
		Mount Royal University, Canada, \texttt{rslawinski@mtroyal.ca}
	}\,,
	Theodore Stanoev%
	\footnote{
		Memorial University of Newfoundland, Canada, \texttt{theodore.stanoev@gmail.com}
	}
}
\date{October 12, 2021}
%%%%%%%%%%%%%%%%%%%%%%%%%
\maketitle
%%%%%%%%%%%%%%%%%%%%%%%%%
\begin{abstract}
%%%%%%%%%%%%%%%%%%%%%%%%%
\noindent We model the instantaneous power applied by a cyclist on a velodrome\,---\,for individual pursuits and other individual time trials\,---\,taking into account its straights, circular arcs, and connecting transition curves.
The forces opposing the motion are air resistance, rolling resistance, lateral friction and drivetrain resistance.
We examine the constant-cadence and constant-power cases, and discuss their results, including an examination of empirical adequacy of the model.
%%%%%%%%%%%%%%%%%%%%%%%%%
\end{abstract}
%%%%%%%%%%%%%%%%%%%%%%%%%
\section{Introduction}
%%%%%%%%%%%%%%%%%%%%%%%%%
In this article, we formulate a mathematical model to examine the power expended by a cyclist on a velodrome.
Our assumptions limit the model to such races as the individual pursuit, a kilometer time trial and the hour record; we do not include drafting effects, which arise in team pursuits.
Furthermore, our model does not account for the initial acceleration; we assume the cyclist to be already in a launched effort.

The only opposing forces we consider are dissipative forces, namely, air resistance, rolling resistance, lateral friction and drivetrain resistance.
We do not consider changes in mechanical energy, which\,---\,on a velodrome\,---\,result from the change of height of the centre of mass due to leaning along the curves.
Nevertheless, our model is empirically adequate \citep{Fraassen}; it accounts for measurements with a satisfactory accuracy.

This article is a continuation of research presented by \citet{DSSbici1}, \citet{DSSbici2}, \citet{BSSbici6} and, in particular, by \citet{SSSbici3} and \citet{BSSSbici4}.
Several details\,---\,for conciseness omitted herein\,---\,are referred to therein.

Geometrically, we consider a velodrome with its straights, circular arcs, and connecting transition curves, whose inclusion\,---\,while presenting a certain challenge, and neglected in previous studies \citep[e.g.,][]{SSSbici3}\,---\,increases the empirical adequacy of the model, as discussed by, among others, \citet{Solarczyk}.

We begin this article by expressing mathematically the geometry of both the black line%
\footnote{The circumference along the inner edge of this five-centimetre-wide line\,---\,also known as the measurement line and the datum line\,---\,corresponds to the official length of the track.}
and the inclination of the track.
Our expressions are accurate representations of the common geometry of modern $250$\,-metre velodromes~(Mehdi Kordi, {\it pers.~comm.}, 2020).
We proceed to formulate an expression for power expended against dissipative forces, which we examine for both the constant-cadence and constant-power cases.
We examine their empirical adequacy, and conclude by discussing the results.
%%%%%%%%%%%%%%%%%%%%%%%%%
\section{Track}
\label{sec:Formulation}
%%%%%%%%%%%%%%%%%%%%%%%%%
\subsection{Black-line parameterization}
\label{sub:Track}
\begin{figure}[h]
\centering
\includegraphics[scale=0.35]{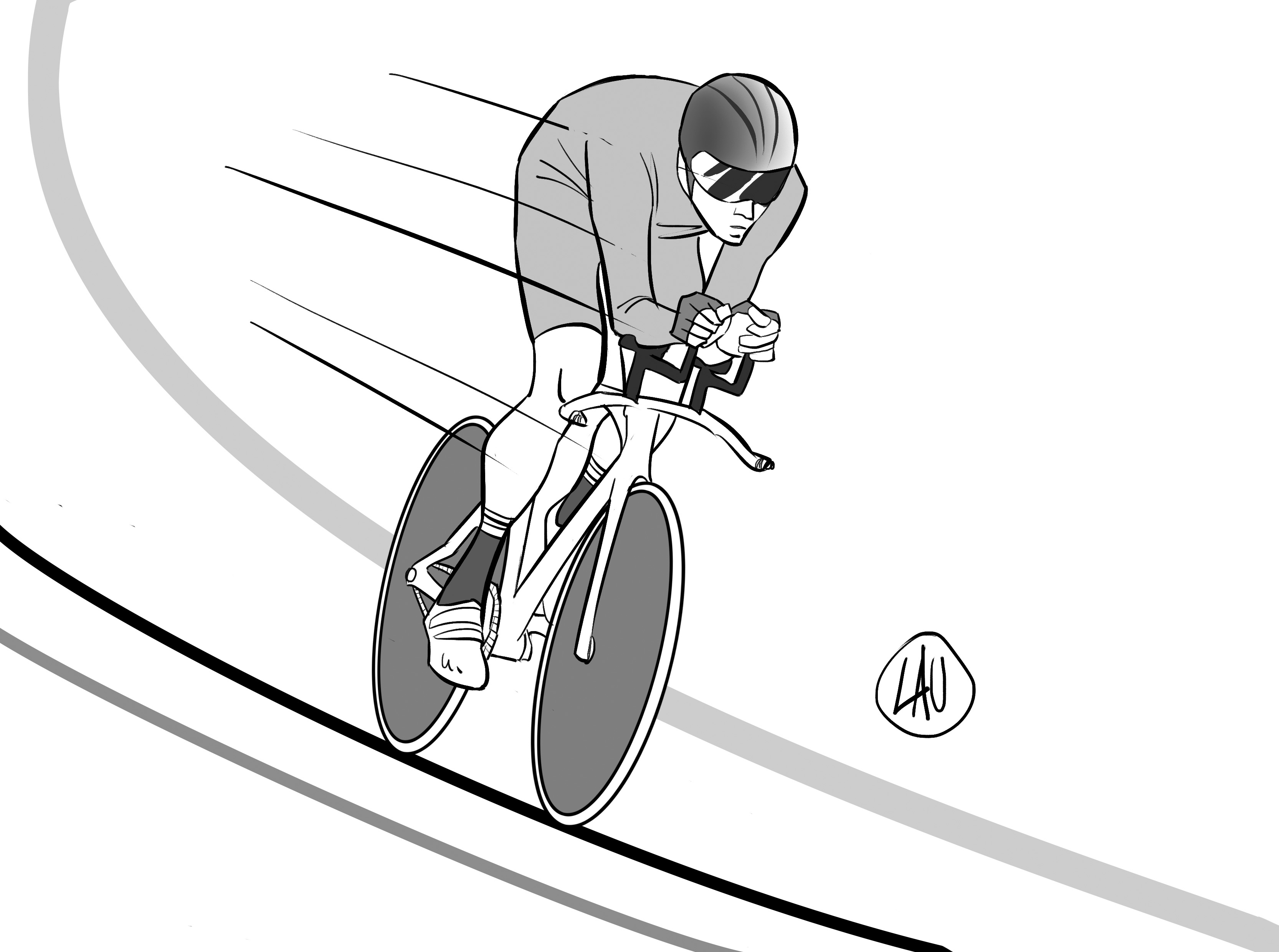}
\caption{\small  A constant aerodynamic position along the black line}
\label{fig:FigBlackLine}
\end{figure}

\begin{figure}[h]
\centering
\includegraphics[scale=0.5]{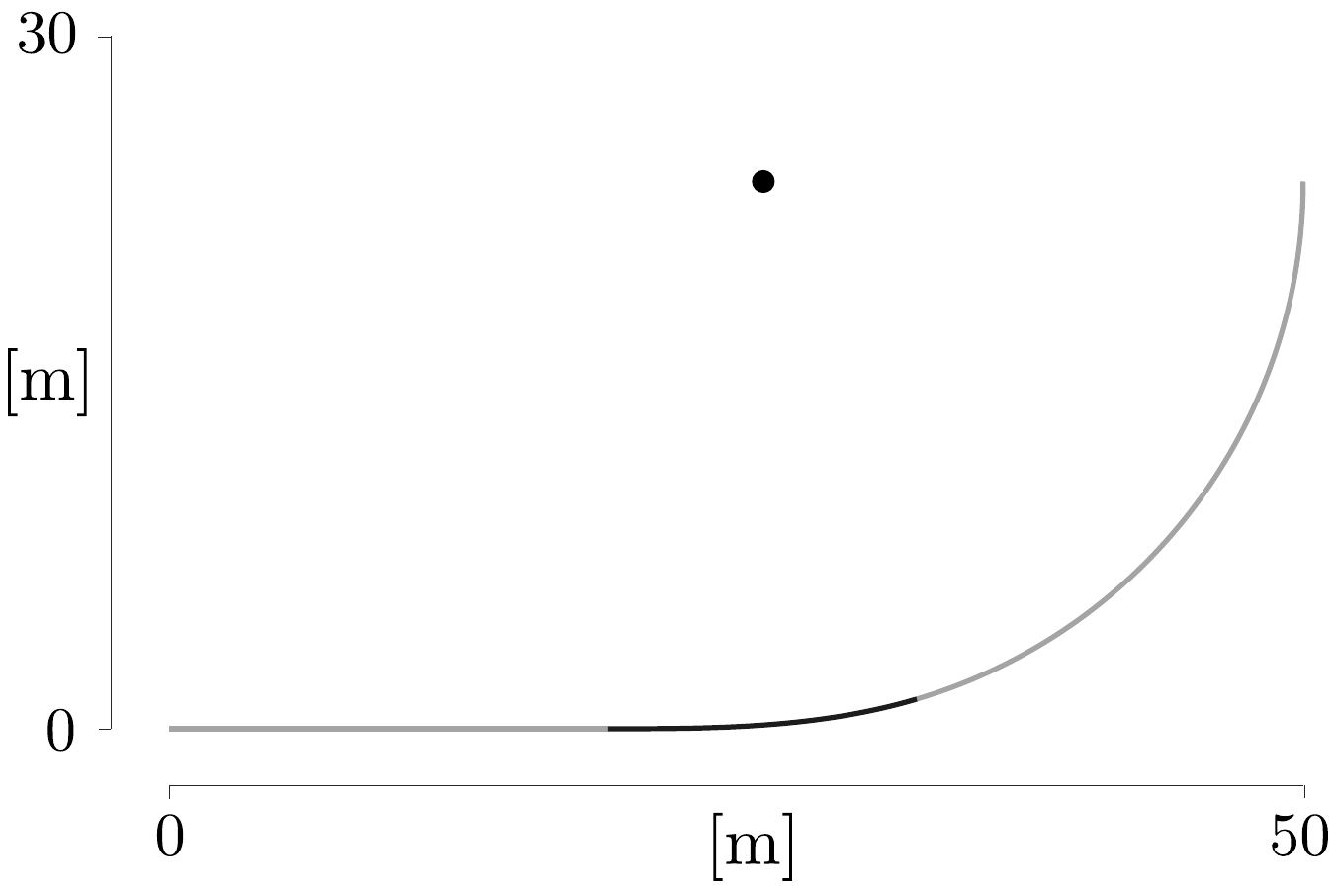}
\caption{\small  A quarter of the black line for a $250$\,-metre track}
\label{fig:FigTrack}
\end{figure}

%%%%%%%%%%%%%%%%%%%%%%%%%
To model the required power of a cyclist who follows the black line, in a constant aerodynamic position, as illustrated in Figure~\ref{fig:FigBlackLine}, we define this line by three parameters.
%%%%%%%%%%%%%%%%%%%%%%%%%
%%%%%%%%%%%%%%%%%%%%%%%%%
\begin{itemize}
\item[--] $L_s$\,: the half-length of the straight
\item[--] $L_t$\,: the length of the transition curve between the straight and the circular arc
\item[--] $L_a$\,: the half-length of the circular arc
\end{itemize}
The length of the track is $S=4(L_s+L_t+L_a)$\,.
In Figure~\ref{fig:FigTrack}, we show a quarter of a black line for $L_s=19$\,m\,, $L_t=13.5$\,m and $L_a=30$\,m\,, which results in $S=250$\,m\,. 
%%%%%%%%%%%%%%%%%%%%%%%%%
%%%%%%%%%%%%%%%%%%%%%%%%%
This curve has continuous derivative up to order two; it is a $C^2$ curve,  whose curvature is continuous.

To formulate, in Cartesian coordinates, the curve shown in Figure~\ref{fig:FigTrack}, we consider the following.
\begin{itemize}
\item[--] The straight,
\begin{equation*}
y_1=0\,,\qquad0\leqslant x\leqslant a\,,	
\end{equation*}
shown in gray, where $a:=L_s$\,.
\item[--] The transition, shown in black\,---\,following a standard design practice\,---\,we take to be an Euler spiral, which can be parameterized by Fresnel integrals,
\begin{equation*}
x_2(\varsigma)=a+\sqrt{\frac{2}{A}}\int\limits_0^{\varsigma\sqrt{\!\frac{A}{2}}}\!\!\!\!\cos\!\left(x^2\right)\,{\rm d}x
\qquad{\rm and}\qquad
y_2(\varsigma)=\sqrt{\frac{2}{A}}\int\limits_0^{\varsigma\sqrt{\!\frac{A}{2}}}\!\!\!\!\sin\!\left(x^2\right)\,{\rm d}x\,,
\end{equation*}
with $A>0$ to be determined; herein, $\varsigma$ is the curve parameter.
Since the arclength differential,~${\rm d}s$\,, is such that
\begin{equation*}
{\rm d}s=\sqrt{x_2'(\varsigma)^2+y_2'(\varsigma)^2}\,{\rm d}\varsigma
=\sqrt{\cos^2\left(\dfrac{A\varsigma^2}{2}\right)+\sin^2\left(\dfrac{A\varsigma^2}{2}\right)}\,{\rm d}\varsigma
={\rm d}\varsigma\,,
\end{equation*}
we write the transition curve as
\begin{equation*}
(x_2(s),y_2(s)), \quad 0\leqslant s\leqslant b:=L_t\,.
\end{equation*}
\item[--] The circular arc, shown in gray, whose centre is $(c_1,c_2)$ and whose radius is $R$\,, with $c_1$\,, $c_2$ and $R$  to be determined.
Since its arclength is specified to be $c:=L_a,$ we may parameterize the quarter circle by
\begin{equation}
\label{eq:x3}
x_3(\theta)=c_1+R\cos(\theta)
\end{equation}
and
\begin{equation}
\label{eq:y3}
y_3(\theta)=c_2+R\sin(\theta)\,,
\end{equation}
where $-\theta_0\leqslant\theta\leqslant 0$\,, for $\theta_0:=c/R$\,.
The centre of the circle is shown as a black dot in Figure~\ref{fig:FigTrack}.
\end{itemize}

We wish to connect these three curve segments so that the resulting global curve is continuous along with its first and second derivatives.
This ensures that the curvature of the track is also continuous.

To do so, let us consider the connection between the straight and the Euler spiral.
Herein, $x_2(0)=a$ and $y_2(0)=0$\,, so the spiral connects continuously to the end of the straight at $(a,0)$\,.
Also, at $(a,0)$\,,
\begin{equation*}
\frac{{\rm d}y}{{\rm d}x}=\frac{y_2'(0)}{x_2'(0)}=\frac{0}{1}=0\,,
\end{equation*}
which matches the derivative of the straight line.
Furthermore, the second derivatives match, since
\begin{equation*}
\frac{{\rm d}^2y}{{\rm d}x^2}=\frac{y''_2(0)x_2'(0)-y'_2(0)x_2''(0)}{(x_2'(0))^2}=0\,,
\end{equation*}
which follows, for any $A>0$\,, from
\begin{equation}
\label{eq:FirstDer}
x_2'(\varsigma)=\cos^2\left(\dfrac{A\,\varsigma^2}{2}\right)\,, \quad y_2'(\varsigma)=\sin^2\left(\dfrac{A\,\varsigma^2}{2}\right)
\end{equation}
and
\begin{equation*}
x_2''(\varsigma)=-A\,\varsigma\sin\left(\dfrac{A\,\varsigma^2}{2}\right)\,, \quad y_2''(\varsigma)=A\,\varsigma\cos\left(\dfrac{A\,\varsigma^2}{2}\right)\,.
\end{equation*}
Let us consider the connection between the Euler spiral and the arc of the circle. 
In order that these connect continuously,
\begin{equation*}
\big(x_2(b),y_2(b)\big)=\big(x_3(-\theta_0),y_3(-\theta_0)\big)\,,
\end{equation*}
we require
\begin{equation}
\label{eq:Cont1}
x_2(b)=c_1+R\cos(\theta_0)\,\,\iff\,\,c_1=x_2(b)-R\cos\!\left(\dfrac{c}{R}\right)
\end{equation}
and
\begin{equation}
\label{eq:Cont2}
y_2(b)=c_2-R\sin(\theta_0)\,\,\iff\,\, c_2=y_2(b)+R\sin\!\left(\dfrac{c}{R}\right)\,.
\end{equation}
For the tangents to connect continuously, we invoke expression~(\ref{eq:FirstDer}) to write
\begin{equation*}
(x_2'(b),y_2'(b))=\left(\cos\left(\dfrac{A\,b^2}{2}\right),\,\sin\left(\dfrac{A\,b^2}{2}\right)\right)\,.
\end{equation*}
Following expressions~(\ref{eq:x3}) and (\ref{eq:y3}), we obtain
\begin{equation*}
\big(x_3'(-\theta_0),y_3'(-\theta_0)\big)=\big(R\sin(\theta_0),R\cos(\theta_0)\big)\,,
\end{equation*}
respectively.
Matching the unit tangent vectors results in
\begin{equation}
\label{eq:tangents}
\cos\left(\dfrac{A\,b^2}{2}\right)=\sin\!\left(\dfrac{c}{R}\right)\,,\quad \sin\left(\dfrac{A\,b^2}{2}\right)=\cos\!\left(\dfrac{c}{R}\right)\,.
\end{equation}
For the second derivative, it is equivalent\,---\,and easier\,---\,to match the curvature.
For the Euler spiral,
\begin{equation*}
\kappa_2(s)=\frac{x_2'(s)y_2''(s)-y_2'(s)x_2''(s)}
{\Big(\big(x_2'(s)\big)^2+\big(y_2'(s)\big)^2\Big)^{\frac{3}{2}}}
=A\,s\cos^2\left(\dfrac{A\,s^2}{2}\right)+A\,s\sin^2\left(\dfrac{A\,s^2}{2}\right)
=A\,s\,,
\end{equation*}
which is indeed the defining characteristic of an Euler spiral: the curvature grows linearly in the arclength.
Hence, to match the curvature of the circle at the connection, we require
\begin{equation*}
A\,b=\frac{1}{R} \,\,\iff\,\,A=\frac{1}{b\,R}\,.
\end{equation*}
Substituting this value of $A$ in equations~(\ref{eq:tangents}), we obtain
\begin{align*}
\cos\!\left(\dfrac{b}{2R}\right)&=\sin\!\left(\dfrac{c}{R}\right)\,,\quad \sin\!\left(\dfrac{b}{2R}\right)=\cos\!\left(\dfrac{c}{R}\right)\\
&\iff\dfrac{b}{2R}=\dfrac{\pi}{2}-\dfrac{c}{R}\\
&\iff R=\frac{b+2c}{\pi}.
\end{align*}
It follows that
\begin{equation*}
A=\frac{1}{b\,R}=\frac{\pi}{b\,(b+2c)}\,;
\end{equation*}
hence, the continuity condition stated in expressions~(\ref{eq:Cont1}) and (\ref{eq:Cont2}) determines the centre of the circle,~$(c_1,c_2)$\,.

\begin{figure}[h]
\centering
\includegraphics[scale=0.5]{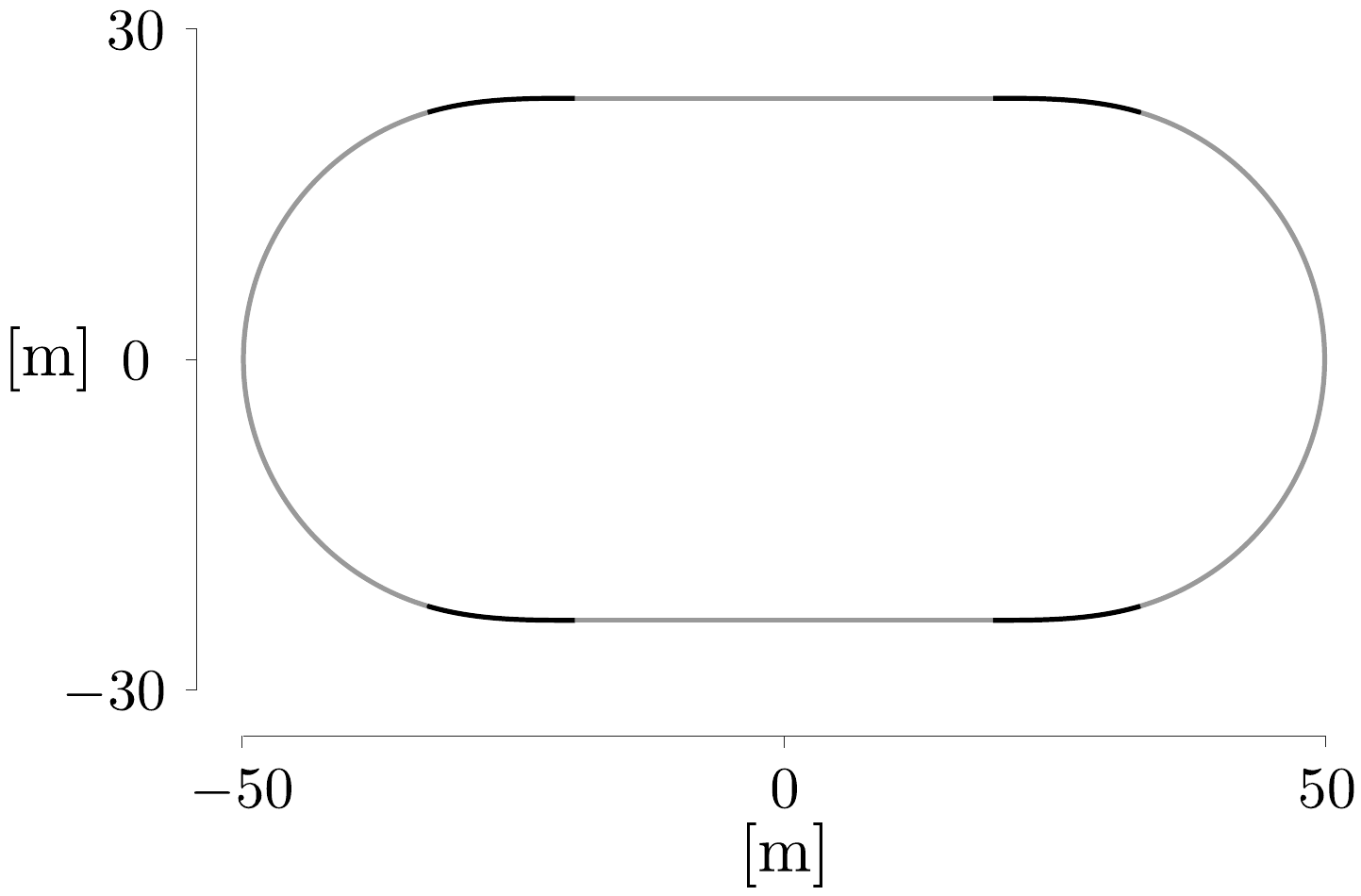}
\caption{\small Black line of $250$\,-metre track}
\label{fig:FigComplete}
\end{figure}

\begin{figure}[h]
\centering
\includegraphics[scale=0.5]{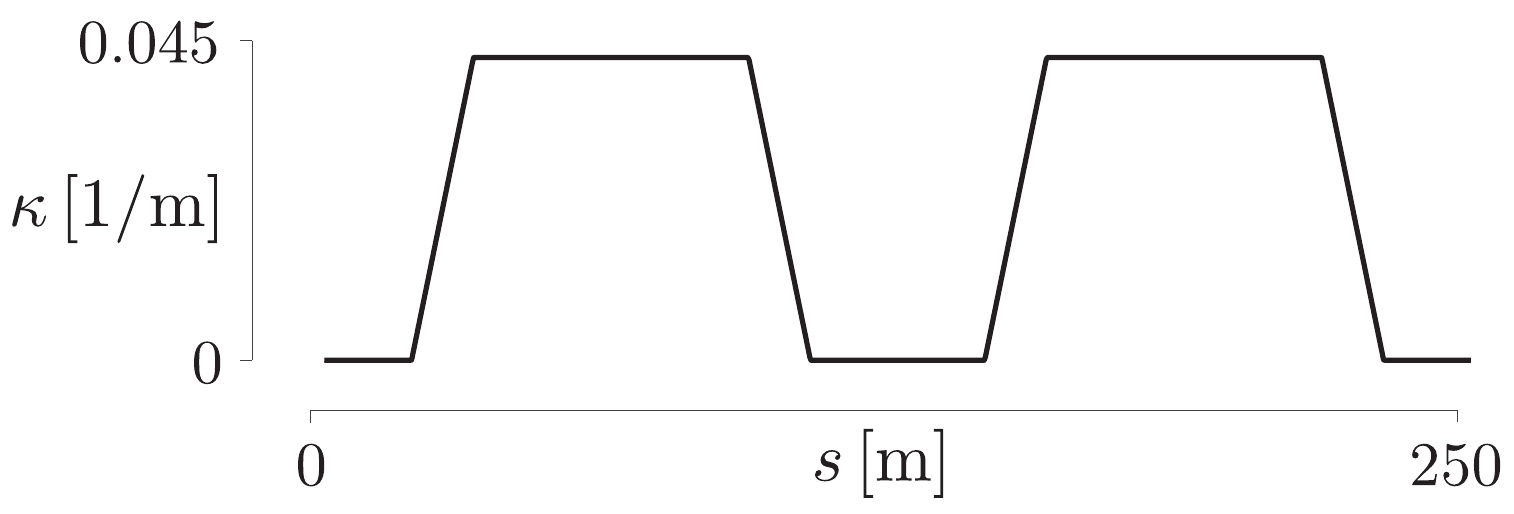}
\caption{\small Curvature of the black line,~$\kappa$\,, as a function of distance,~$s$\,, with a linear transition between the zero curvature of the straight and the $1/R$ curvature of the circular arc}
\label{fig:FigCurvature}
\end{figure}

For the case shown in Figure~\ref{fig:FigTrack}, the numerical values are~$A=3.1661\times10^{-3}$\,m${}^{-2}$, $R=23.3958$\,m\,, $c_1=25.7313$\,m and $c_2=23.7194$\,m\,.
The complete track\,---\,with its centre at the origin\,,~$(0,0)$\,---\,is shown in Figure~\ref{fig:FigComplete}.
%%%%%%%%%%%%%%%%%%%%%%%%%
%%%%%%%%%%%%%%%%%%%%%%%%%
The corresponding curvature is shown in Figure~\ref{fig:FigCurvature}.
Note that  the curvature transitions linearly from the constant value of straight,~$\kappa=0$\,, to the constant value of the circular arc,~$\kappa=1/R$\,.
%%%%%%%%%%%%%%%%%%%%%%%%%
%%%%%%%%%%%%%%%%%%%%%%%%%
%%%%%%%%%%%%%%%%%%%%%%%%%
\subsection{Track-inclination angle}
%%%%%%%%%%%%%%%%%%%%%%%%%
%%%%%%%%%%%%%%%%%%%%%%%%%
\begin{figure}[h]
\centering
\includegraphics[scale=0.5]{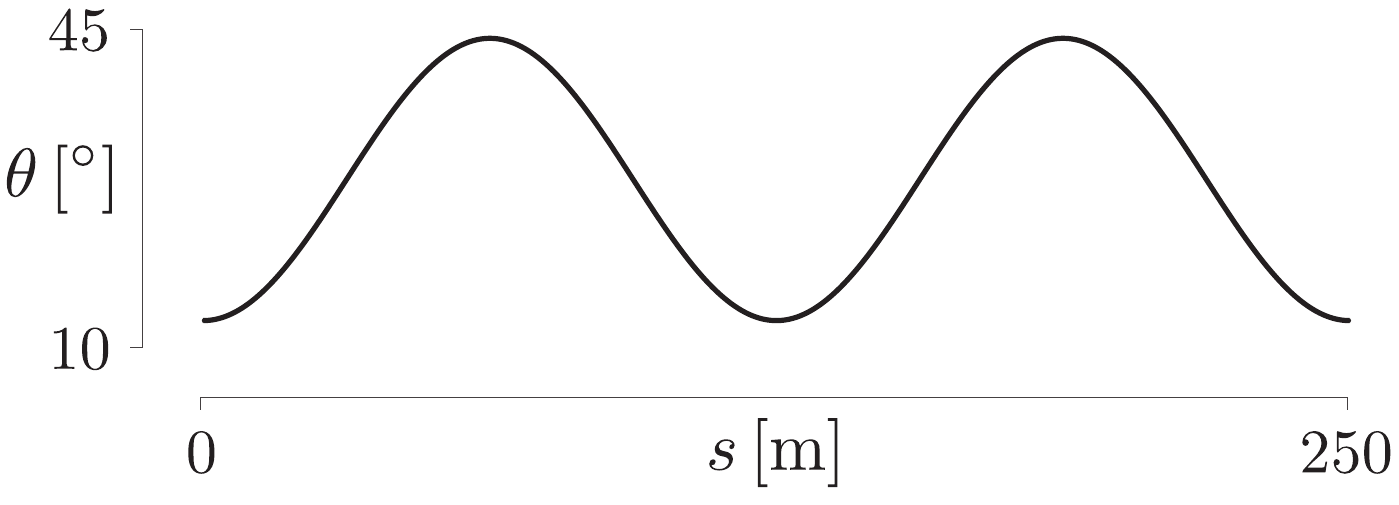}
\caption{\small Track inclination,~$\theta$\,, as a function of the black-line distance,~$s$}
\label{fig:FigAngle}
\end{figure}
%%%%%%%%%%%%%%%%%%%%%%%%%
There are many possibilities to model the track inclination angle.
We choose a trigonometric formula in terms of arclength, which is a good analogy of an actual $250$\,-metre velodrome.
The minimum inclination of $13^\circ$ corresponds to the midpoint of the straight, and the maximum of $44^\circ$ to the apex of the circular arc.
For a track of length $S$\,,
\begin{equation}
\label{eq:theta}
\theta(s)=28.5-15.5\cos\!\left(\frac{4\pi}{S}s\right)\,;
\end{equation}
$s=0$ refers to the midpoint of the lower straight, in Figure~\ref{fig:FigComplete}, and the track is oriented in the counterclockwise direction.
Figure \ref{fig:FigAngle} shows this inclination for $S=250$\,m\,.
\begin{remark}
It is not uncommon for tracks to be slightly asymmetric with respect to the inclination angle.
In such a case, they exhibit a rotational symmetry by $\pi$\,, but not a reflection symmetry about the vertical or horizontal axis.
This asymmetry can be modeled by including $s_0$ in the argument of the cosine in expression~(\ref{eq:theta}).
\begin{equation*}
\theta(s)=28.5-15.5\cos\!\left(\frac{4\pi}{S}(s-s_0)\right)\,;
\end{equation*}
$s_0\approx 5$ provides a good model for several existing velodromes.
Referring to discussions about the London Velodrome of the 2012 Olympics, \citet{Solarczyk} writes that
\begin{quote}
the slope of the track going into and out of the turns is not the same.
This is simply because you always cycle the same way around the track, and you go shallower into the turn and steeper out of it.
\end{quote}
\end{remark}
%%%%%%%%%%%%%%%%%%%%%%%%%
\section{Instantaneous power}
\label{sec:InstPower}
\begin{figure}[h]
\centering
\includegraphics[scale=0.8]{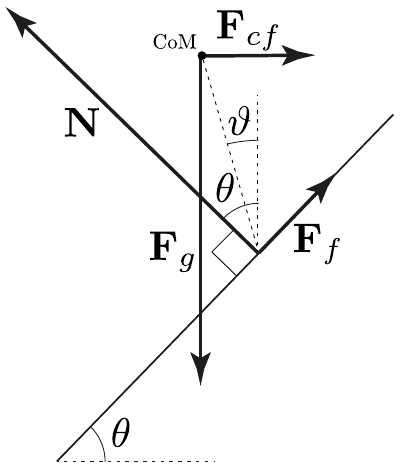}
\caption{\small Force diagram}
\label{fig:FigCentFric}
\end{figure}

%%%%%%%%%%%%%%%%%%%%%%%%%
A mathematical model to account for the power required to propel a bicycle is based on \citep[e.g.,][]{DSSbici1}
\begin{equation}
\label{eq:BikePower}
P=F\,V\,,
\end{equation}
where $F$ stands for the magnitude of forces opposing the motion and $V$ for speed.
Herein, we model the rider as undergoing instantaneous circular motion, in rotational equilibrium about the line of contact of the tires with the ground.
Following \citet[Section~2]{SSSbici3} and in view of Figure~\ref{fig:FigCentFric}, along the black line, in windless conditions,
\begin{subequations}
\label{eq:power}
\begin{align}
\nonumber P&=\\
&\dfrac{1}{1-\lambda}\,\,\Bigg\{\label{eq:modelO}\\
&\left.\left.\Bigg({\rm C_{rr}}\underbrace{\overbrace{\,m\,g\,}^{F_g}(\sin\theta\tan\vartheta+\cos\theta)}_N\cos\theta
+{\rm C_{sr}}\Bigg|\underbrace{\overbrace{\,m\,g\,}^{F_g}\frac{\sin(\theta-\vartheta)}{\cos\vartheta}}_{F_f}\Bigg|\sin\theta\Bigg)\,v
\right.\right.\label{eq:modelB}\\
&+\,\,\tfrac{1}{2}\,{\rm C_{d}A}\,\rho\,V^3\Bigg\}\label{eq:modelC}\,,
\end{align}	
\end{subequations}
where $m$ is the mass of the cyclist and the bicycle, $g$ is the acceleration due to gravity, $\theta$ is the track-inclination angle, $\vartheta$ is the bicycle-cyclist lean angle, $\rm C_{rr}$ is the rolling-resistance coefficient, $\rm C_{sr}$ is the coefficient of the lateral friction, $\rm C_{d}A$ is the air-resistance coefficient, $\rho$ is the air density, $\lambda$ is the drivetrain-resistance coefficient.
Herein, $v$ is the speed at which the contact point of the rotating wheels moves along the track \citep[Appendix~B]{DSSbici1}, which we assume to coincide with the black-line speed.
$V$ is the centre-of-mass speed.
Since lateral friction is a dissipative force, it does negative work, and the work done against it\,---\,as well as the power\,---\,are positive.
For this reason, in expression~(\ref{eq:modelB}), we consider the magnitude,~$\big|{\,\,}\big|$\,.
%%%%%%%%%%%%%%%%%%%%%%%%%
%%%%%%%%%%%%%%%%%%%%%%%%%

To accommodate the assumption of {\it instantaneous} power in the context of measurements within a fixed-wheel drivetrain, as discussed by \citet[Appendix~B.2]{DSSbici2},%
\footnote{For a fixed-wheel drivetrain, the momentum of a bicycle-cyclist system results in rotation of pedals even without any force applied by a cyclist.
This leads to inaccuracy of measurements referring to the instantaneous power generated by a cyclist, which increases with the variability of effort.}
we assume the steadiness of effort, which\,---\,following the initial acceleration\,---\,is consistent with a steady pace of an individual pursuit, as presented in Section~\ref{sec:Adequacy}, below.
Formally, this assumption corresponds to setting the acceleration,~$a$\,, to zero in \citet[expression~(1)]{SSSbici3}, which entails expression~(\ref{eq:power}).
For the constant-cadence case\,---\,in accordance with expression~(\ref{eq:vV}), below\,---\,change of speed refers only to the change of the centre-of-mass speeed; the black-line speed is constant.
For the constant-power case, neither speed is constant, but both are nearly so.

Let us return to expression~(\ref{eq:power}).
Therein, $\theta$ is given by expression~(\ref{eq:theta}).
The lean angle is \citep[Appendix~A]{SSSbici3}
\begin{equation}
\label{eq:LeanAngle}
\vartheta=\arctan\dfrac{V^2}{g\,r_{\rm\scriptscriptstyle CoM}}\,,
\end{equation}
where $r_{\rm\scriptscriptstyle CoM}$ is the centre-of-mass radius, and\,---\,along the curves, at any instant\,---\,the centre-of-mass speed is
\begin{equation}
\label{eq:vV}
V=v\,\dfrac{\overbrace{(R-h\sin\vartheta)}^{\displaystyle r_{\rm\scriptscriptstyle CoM}}}{R}
=v\,\left(1-\dfrac{h\,\sin\vartheta}{R}\right)\,,
\end{equation}
where $R$ is the radius discussed in Section~\ref{sub:Track} and $h$ is the centre-of-mass height of the bicycle-cyclist system at $\vartheta=0$\,.
Along the straights, the  black-line speed is equal to the centre-of-mass speed, $v=V$\,.
As expected, $V=v$ if $h=0$\,, $\vartheta=0$ or $R=\infty$\,.

As illustrated in Figure~\ref{fig:FigCentFric}, expressions~(\ref{eq:LeanAngle}) and (\ref{eq:vV}) assume instantaneous circular motion of the centre of mass to occur in a horizontal plane.
Therefore, using these expression implies neglecting the vertical motion of the centre of mass.
Accounting for the vertical motion of the centre of mass would mean allowing for a nonhorizontal centripetal force, and including the work done in raising the centre of mass.
%%%%%%%%%%%%%%%%%%%%%%%%%
\section{Numerical examples}
\label{sec:NumEx}
%%%%%%%%%%%%%%%%%%%%%%%%%
\subsection{Model-parameter values}
\label{sub:ModPar}
%%%%%%%%%%%%%%%%%%%%%%%%%
For expressions~(\ref{eq:power}), (\ref{eq:LeanAngle}) and (\ref{eq:vV}), we consider a velodrome discussed in Section~\ref{sec:Formulation}, and let ${R=23.3958}$\,m\,.
For the bicycle-cyclist system, we assume, $h=1.2$\,m\,, $m=84$\,kg\,, ${\rm C_{d}A}=0.2$\,m${}^2$\,, ${\rm C_{rr}}=0.002$\,, ${\rm C_{sr}}=0.003$ and $\lambda=0.02$\,.
For the external conditions, $g=9.81$\,m/s${}^2$ and $\rho=1.225$\,kg/m${}^3$\,.
%%%%%%%%%%%%%%%%%%%%%%%%%
\subsection{Constant cadence}
\label{sub:ConstCad}
\begin{figure}[h]
\centering
\includegraphics[scale=0.5]{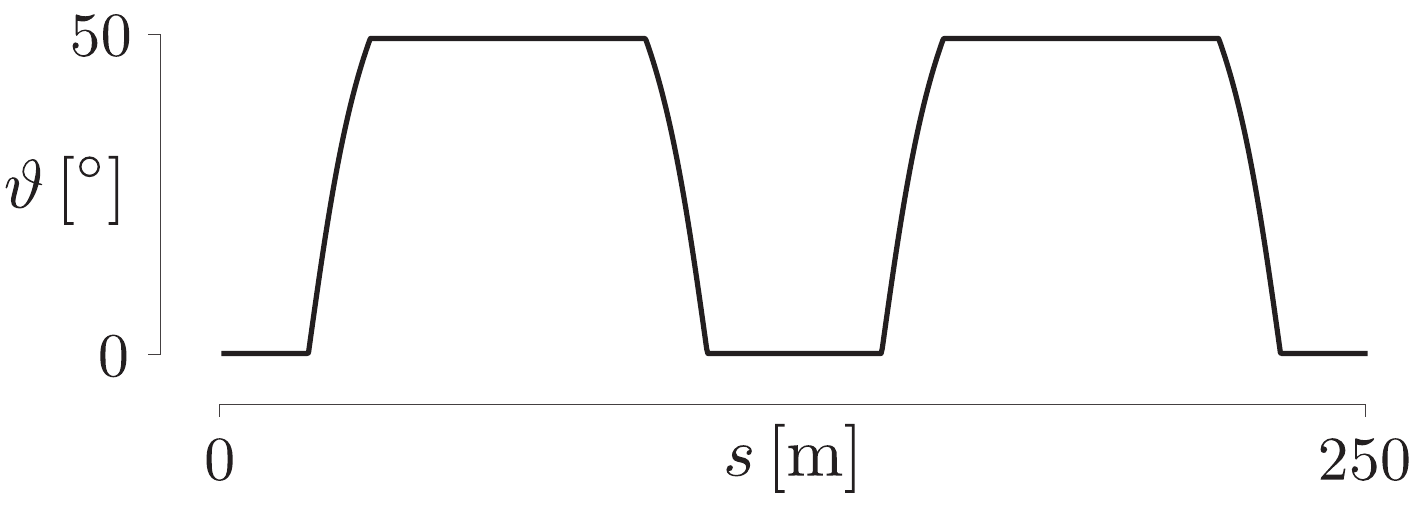}
\caption{\small Lean angle,~$\vartheta$\,, as a function of the black-line distance,~$s$\,, for constant cadence}
\label{fig:FigLeanAngle}
\end{figure}

\begin{figure}[h]
\centering
\includegraphics[scale=0.5]{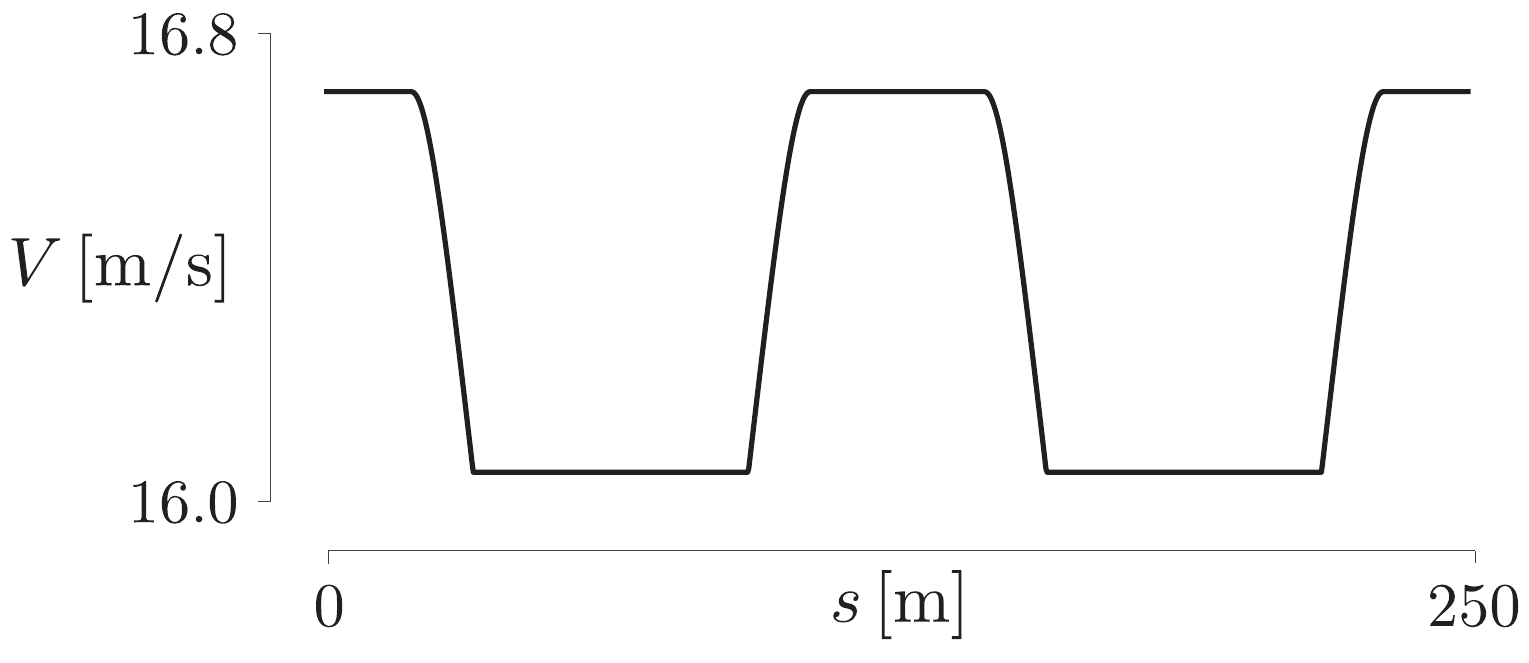}
\caption{\small Centre-of-mass speed,~$V$\,, as a function of the black-line distance,~$s$\,, for constant cadence}
\label{fig:FigCoMSpeed}
\end{figure}

%%%%%%%%%%%%%%%%%%%%%%%%%
Let the black-line speed be constant,~$v=16.7$\,m/s\,, which is tantamount to the constancy of cadence.
The lean angle and the centre-of-mass speed, as functions of distance\,---\,obtained by numerically and simultaneously solving equations~(\ref{eq:LeanAngle}) and (\ref{eq:vV}), at each point of a discretized model of  the track\,---\,are shown in Figures~\ref{fig:FigLeanAngle} and \ref{fig:FigCoMSpeed}, respectively.
The average centre-of-mass speed, per lap is~$\overline V=16.3329$\,m/s\,.
Changes in $V$\,, shown in Figure~\ref{fig:FigCoMSpeed}, result from changes in the lean angle.
Along the straights, $\vartheta=0\implies V=v$\,.
Along the curves, since $\vartheta\neq0$\,, the centre-of-mass travels along a shorter path; hence, $V<v$\,.
Thus, assuming a constant black-line speed implies a variable centre-of-mass speed and, hence, an acceleration and deceleration, even though ${\rm d}V/{\rm d}t$\,, where $t$ stands for time, is not included explicitly in expression~(\ref{eq:power}).
Examining Figure~\ref{fig:FigCoMSpeed}, we conclude that ${\rm d}V/{\rm d}t\neq0$ along the transition curves only.

\begin{figure}[h]
\centering
\includegraphics[scale=0.5]{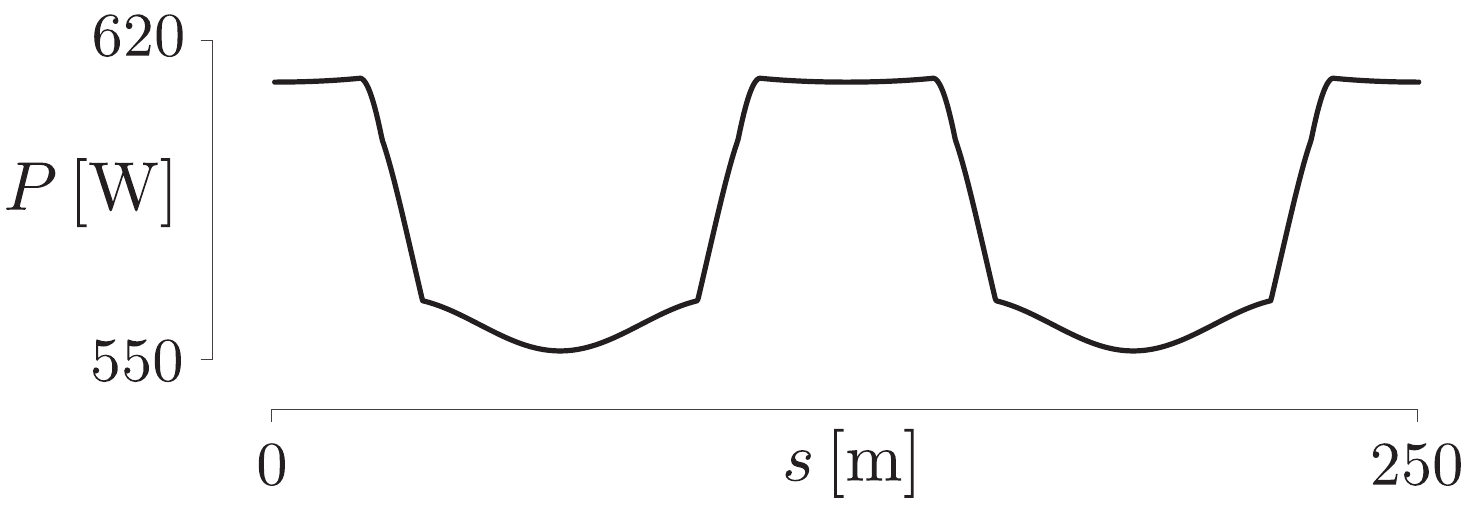}
\caption{\small Power,~$P$\,, as a function of the black-line distance,~$s$\,, for constant cadence}
\label{fig:FigPower}
\end{figure}

The power\,---\,obtained by evaluating expression~(\ref{eq:power}), at each point along the track\,---\,is shown in Figure~\ref{fig:FigPower}.
The average power, per lap, is $\overline P=580.5941$\,W\,.
Since the black-line speed is constant, this is both the arclength average and the temporal average.
%%%%%%%%%%%%%%%%%%%%%%%%%
%%%%%%%%%%%%%%%%%%%%%%%%%
%%%%%%%%%%%%%%%%%%%%%%%%%
%%%%%%%%%%%%%%%%%%%%%%%%%
%%%%%%%%%%%%%%%%%%%%%%%%%
%%%%%%%%%%%%%%%%%%%%%%%%%
%%%%%%%%%%%%%%%%%%%%%%%%%
\begin{figure}[h]
\centering
\includegraphics[scale=0.5]{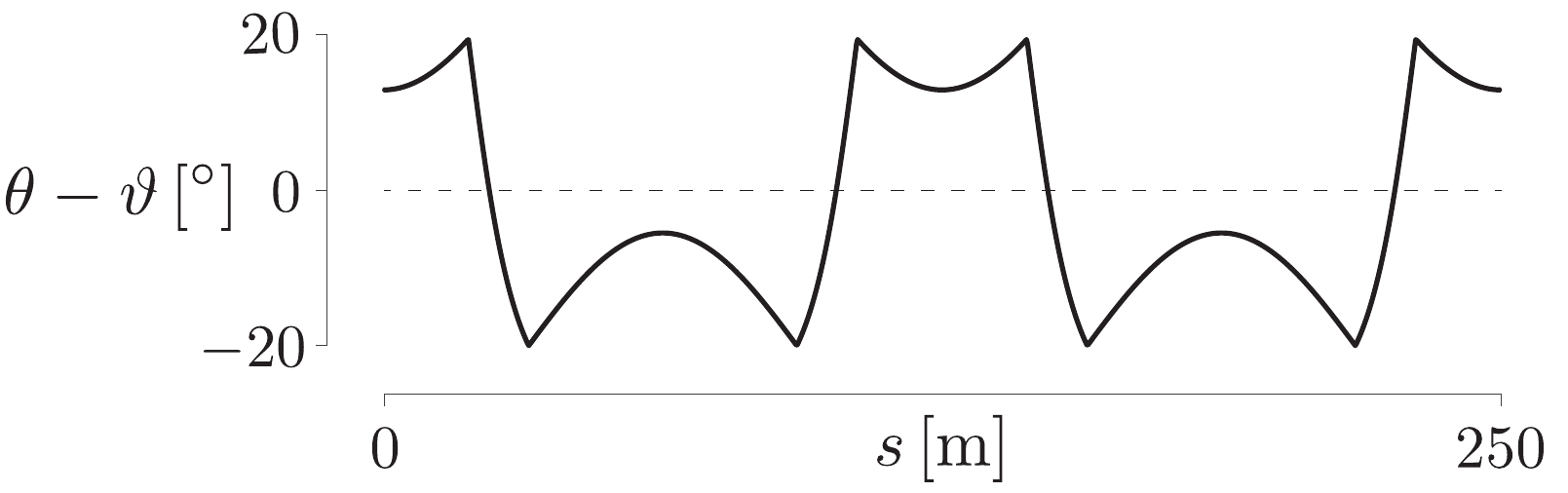}
\caption{\small $\theta-\vartheta$\,, as a function of the black-line distance,~$s$\,, for constant cadence}
\label{fig:FigAngleDiff}
\end{figure}
%%%%%%%%%%%%%%%%%%%%%%%%%
%%%%%%%%%%%%%%%%%%%%%%%%%
\begin{figure}[h]
\centering
\includegraphics[scale=0.5]{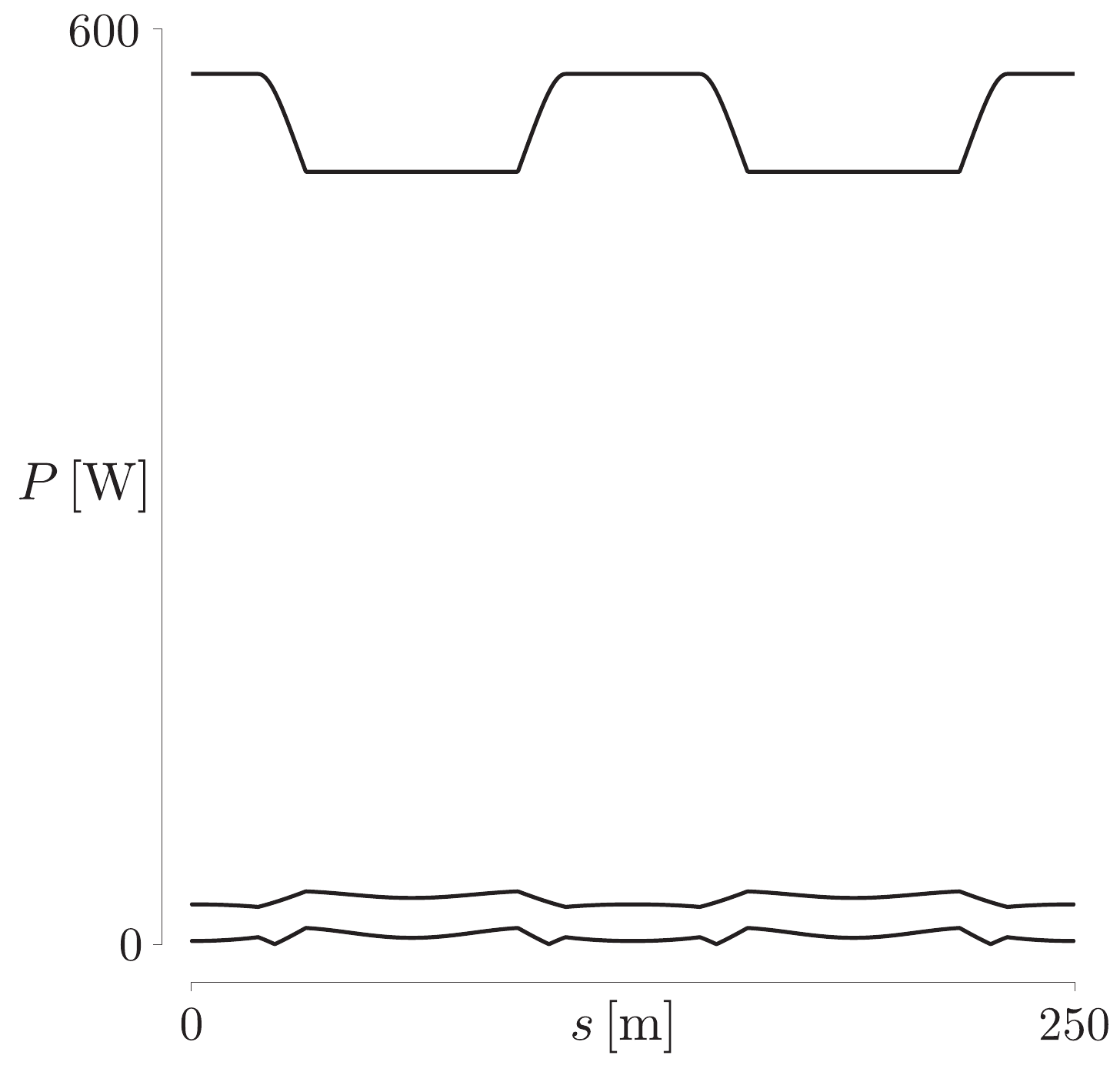}
\caption{\small Power to overcome air resistance, rolling resistance and lateral friction}
\label{fig:FigPowerSummands}
\end{figure}
%%%%%%%%%%%%%%%%%%%%%%%%%

Examining Figure~\ref{fig:FigPower}, we see the decrease of power along the curve to maintain the same black-line speed.
This is due to both the decrease of the centre-of-mass speed, which results in a smaller value of term~(\ref{eq:modelC}), and the decrease of a difference between the track-inclination angle and the lean angle, shown in Figure~\ref{fig:FigAngleDiff}, which results in a smaller value of the second summand of term~(\ref{eq:modelB}).

Examining Figure~\ref{fig:FigPowerSummands}, where\,---\,in accordance with expression~(\ref{eq:power})\,---\,we distinguish among the power used to overcome the air resistance, the rolling resistance and the lateral friction, we can quantify their effects.
The first has the most effect; the last has the least effect, and is zero at points for which $\theta=\vartheta$\,, which corresponds to the zero crossings in Figure~\ref{fig:FigAngleDiff}.

Let us comment on potential simplifications of the model.
If we assume a straight flat course\,---\,which is tantamount to neglecting the lean and inclination angles\,---\,we obtain $\overline P\approx 610~\rm W$\,.
If we consider an oval track but ignore the transitions and assume that the straights are flat and the semicircular segments, whose radius is $23$\,m\,, have a constant inclination of $43^\circ$, we obtain \citep[expression~(13)]{SSSbici3} $\overline P\approx 563~\rm W$\,.
In both cases, there is a significant discrepancy with the power obtained from the model discussed herein,~$\overline P=580.5941~\rm W$\,.

To conclude this section, let us calculate the work per lap corresponding to the model discussed herein.
The work performed during a time interval, $t_2-t_1$\,, is
\begin{equation*}
W=\int\limits_{t_1}^{t_2}\!P\,{\rm d}t
=\dfrac{1}{v}\int\limits_{s_1}^{s_2}\!P\!\underbrace{\,v\,{\rm d}t}_{{\rm d}s\,}\,,	
\end{equation*}
where the black-line speed,~$v$\,, is constant and, hence, ${\rm d}s$ is an arclength distance along the black line.
Considering the average power per lap, we write
\begin{equation}
\label{eq:WorkConstCad}
W=\underbrace{\,\dfrac{S}{v}\,}_{t_\circlearrowleft}\,\underbrace{\dfrac{\int\limits_0^S\!P\,{\rm d}s}{S}}_{\overline P}=\overline P\,t_\circlearrowleft\,.
\end{equation}
Given $\overline P=580.5941$\,W and $t_\circlearrowleft=14.9701$\,s\,, we obtain $W=8691.5284$\,J\,.
%%%%%%%%%%%%%%%%%%%%%%%%%
\subsection{Constant power}
\label{sub:ConstPower}
%%%%%%%%%%%%%%%%%%%%%%%%%
Let us solve numerically the system of nonlinear equations given by  expressions~(\ref{eq:power}),  (\ref{eq:LeanAngle}) and (\ref{eq:vV}), to find the lean angle as well as both speeds, $v$ and $V$\,, at each point of a discretized model of the track\,, under the assumption of constant power.
As in Section~\ref{sub:ConstCad}, we let $R=23.3958$\,m\,, $h=1.2$\,m\,, $m=84$\,kg\,, ${\rm C_{d}A}=0.2$\,m${}^2$\,, ${\rm C_{rr}}=0.002$\,, ${\rm C_{sr}}=0.003$\,, $\lambda=0.02$\,, $g=9.81$\,m/s${}^2$ and $\rho=1.225$\,kg/m${}^3$\,.
However, in contrast to Section~\ref{sub:ConstCad}, we allow the black-line speed to vary but set the power to be the average obtained in that section, $P=580.5941$\,W\,.

Stating expression~(\ref{eq:vV}), as
\begin{equation}
\label{eq:Vv}
v=V\dfrac{R}{R-h\sin\vartheta}\,,
\end{equation}
we write expression~(\ref{eq:power}) as
\begin{align}
\label{eq:PConst}
P&=\\
\nonumber&\dfrac{V}{1-\lambda}\,\,\Bigg\{\\
\nonumber&\left.\left.\Bigg({\rm C_{rr}}\,m\,g\,(\sin\theta\tan\vartheta+\cos\theta)\cos\theta
+{\rm C_{sr}}\Bigg|\,m\,g\,\frac{\sin(\theta-\vartheta)}{\cos\vartheta}\Bigg|\sin\theta\Bigg)\,\dfrac{R}{R-h\sin\vartheta}
\right.\right.\\
\nonumber&+\,\,\tfrac{1}{2}\,{\rm C_{d}A}\,\rho\,V^2\Bigg\}\,,
\end{align}
and expression~(\ref{eq:LeanAngle}) as
\begin{equation}
\label{eq:Vvar}
\vartheta=\arctan\dfrac{V^2}{g\,(R-h\sin\vartheta)}\,,
\end{equation}
which\,---\,given $g$\,, $R$ and $h$\,---\,can be solved for $V$ as a function of~$\vartheta$\,.
Inserting that solution in expression~(\ref{eq:PConst}), we obtain an equation whose only unknown is~$\vartheta$\,.

The difference of the lean angle\,---\,between the case of a constant cadence and a constant power\,---\,is so small that there is no need to plot it; Figure~\ref{fig:FigLeanAngle} illustrates it accurately.
The same is true for the difference between the track-inclination angle and the lean angle, illustrated in Figure~\ref{fig:FigAngleDiff}, as well as for the dominant effect of the air resistance, illustrated in Figure~\ref{fig:FigPowerSummands}.

\begin{figure}[h]
\centering
\includegraphics[scale=0.5]{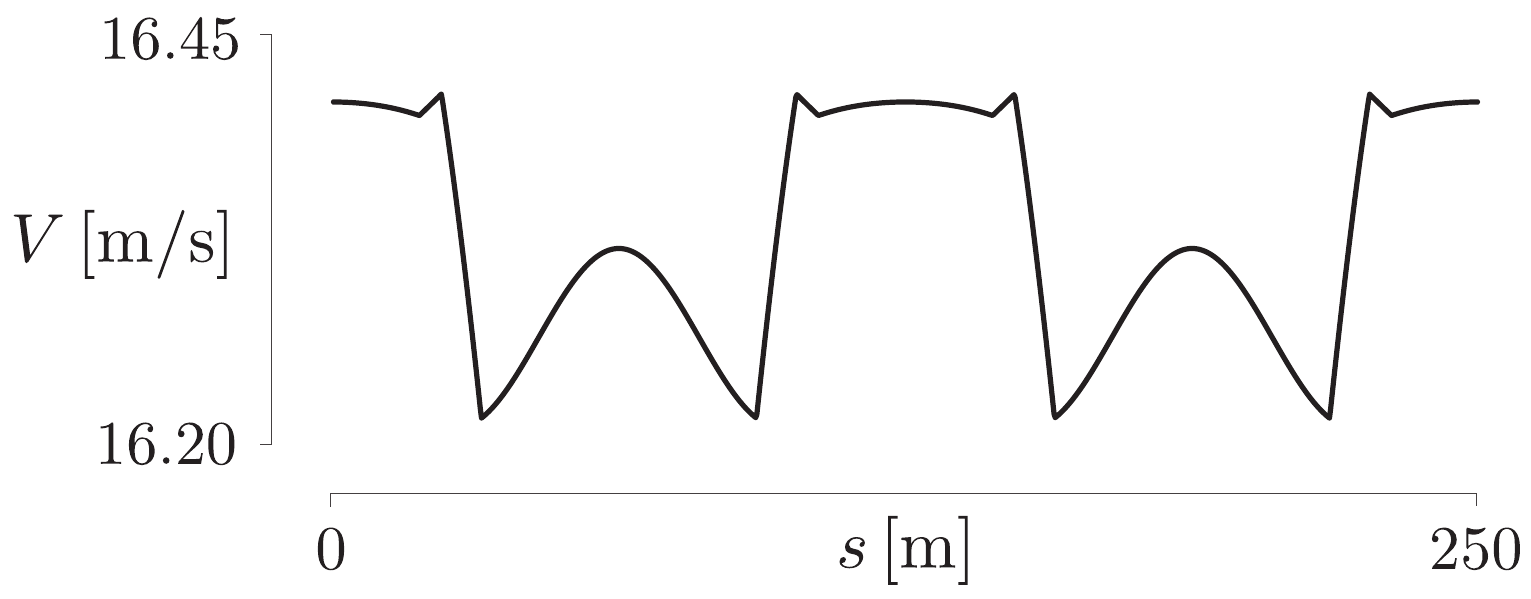}
\caption{\small Centre-of-mass speed,~$V$\,, as a function of the black-line distance,~$s$\,, for constant power}
\label{fig:FigCoMSpeed2}
\end{figure}

The resulting values of $V$ are shown in Figure~\ref{fig:FigCoMSpeed2}.
As expected, in view of the dominant effect of air resistance, a constancy of $P$ entails only small variations in~$V$\,.
In comparison to the case discussed in Section~\ref{sub:ConstCad}, the case in question entails lesser changes of the centre-of-mass speed\,---\,note the difference of vertical scale between Figures~\ref{fig:FigCoMSpeed} and \ref{fig:FigCoMSpeed2}\,---\,but the changes of speed are not limited to the transition curves.
Even though such changes are not included explicitly in expression~(\ref{eq:power}), a portion of the given power is due to the $m\,V\,{\rm d}V/{\rm d}t$ term, which is associated with acceleration and deceleration.
The amount of this portion can be estimated {\it a posteriori}.

Since
\begin{equation}
\label{eq:dK}
m\,V\,\dfrac{{\rm d}V}{{\rm d}t}=\dfrac{{\rm d}}{{\rm d}t}\left(\dfrac{1}{2}\,m\,V^2\right)\,,
\end{equation}
the time integral of the power used for acceleration of the centre of mass is the change of its kinetic energy.
Therefore, to include the effect of accelerations, per lap, we need to add the increases in kinetic energy.
This is an estimate of the error committed by neglecting accelerations in expression~(\ref{eq:power}), which could be quantified, for the constant-cadence and constant-power cases, following \citet[Appendix~A]{BSSSbici4}.
Also, in the same appendix, \citeauthor{BSSSbici4} discuss errors due to neglecting raising the centre of mass upon exiting the curve, which is an increase of potential energy.
As mentioned in Section~\ref{sec:InstPower}, herein, expressions~(\ref{eq:LeanAngle}) and (\ref{eq:vV}) assume instantaneous circular motion of the centre of mass to occur in a horizontal plane.

For inverse problems, the power used to increase the kinetic and potential energy is implicitly included on the left-hand side of  expression~(\ref{eq:power}).
Hence, the power required to increase mechanical energy is incorporated in the power to account for the drivetrain, tire and air frictions stated in expressions~(\ref{eq:modelO}), (\ref{eq:modelB}) and (\ref{eq:modelC}), respectively.
Since the model does not explicitly incorporate changes in mechanical energy, this necessarily leads to an overestimate of the values of $\lambda$\,, $\rm C_{rr}$\,, $\rm C_{sr}$ and $\rm C_{d}A$ to achieve the agreement between the model and measurements.
A model that explicitly takes into account the acceleration and vertical motion of the centre of mass is to be considered in future work.

\begin{figure}[h]
\centering
\includegraphics[scale=0.5]{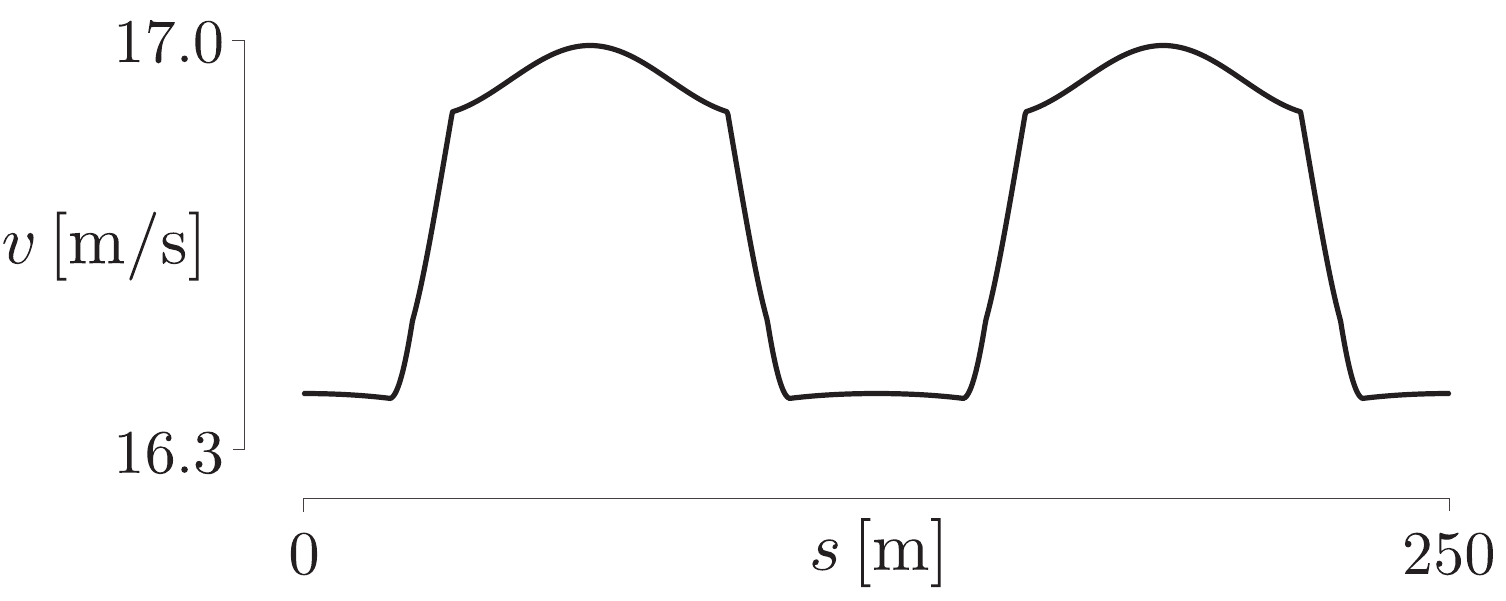}
\caption{\small Black-line speed,~$v$\,, as a function of the black-line distance,~$s$\,, for constant power}
\label{fig:FigBLspeed}
\end{figure}

The values of $v$\,, in accordance with expression~(\ref{eq:vV}), are shown in Figure~\ref{fig:FigBLspeed}, where\,---\,as expected for a constant power\,---\,leaning into the turn entails an increase of the black-line speed; note the difference of vertical scale between Figures~\ref{fig:FigCoMSpeed2} and \ref{fig:FigBLspeed}.
The averages are $\overline V=16.3316$\,m/s and $\overline v=16.7071$\,m/s\,.
These averages are similar to the case of the constant black-line speed averages.
Hence, maintaining a constant cadence or a constant power results in nearly the same laptime, namely, $14.9701$\,s and $14.9670$\,s\,, respectively.

To conclude this section, let us calculate the corresponding work per lap.
The work performed during a time interval, $t_2-t_1$\,, is
\begin{equation}
\label{eq:WorkConstPow}
W=\int\limits_{t_1}^{t_2}\!P\,{\rm d}t=P\!\int\limits_{t_1}^{t_2}\!{\rm d}t=P\,\underbrace{(t_2-t_1)}_{t_\circlearrowleft}=P\,t_\circlearrowleft\,,	
\end{equation}
where, for the second equality sign, we use the constancy of~$P$\,; also, we let the time interval to be a laptime.
Thus, given $\overline P=580.5941$\,W and $t_\circlearrowleft=14.9670$\,s\,, we obtain $W=8689.7680$\,J\,.

%%%%%%%%%%%%%%%%%%%%%%%%%
%%%%%%%%%%%%%%%%%%%%%%%%%
%%%%%%%%%%%%%%%%%%%%%%%%%
%%%%%%%%%%%%%%%%%%%%%%%%%
The empirical adequacy of the assumption of a constant power can be corroborated\,---\,apart from the measured power itself\,---\,by comparing experimental data to measurable quantities entailed by theoretical formulations.
The black-line speed,~$v$\,, shown in Figure~\ref{fig:FigBLspeed}, which we take to be tantamount to the wheel speed, appears to be the most reliable quantity.
Notably, an increase of $v$ by a few percent along the turns is a commonly measured quantity.
Other quantities\,---\,not measurably directly, such as the centre-of-mass speed and power expended to increase potential energy\,---\,are related to $v$ by equations~(\ref{eq:PConst}) and (\ref{eq:Vvar}).
%%%%%%%%%%%%%%%%%%%%%%%%%
\section{Empirical adequacy}
\label{sec:Adequacy}
%%%%%%%%%%%%%%%%%%%%%%%%%
To gain an insight into empirical adequacy of the model, let us examine Section~\ref{sec:InstPower} in the context of measurements~(Mehdi Kordi, {\it pers.~comm.}, 2020).
To do so, we use two measured quantities: cadence and force applied to the pedals, both of  which are measured by sensors attached to the bicycle.
They allow us to calculate power, which is the product of the circumferential pedal speed\,---\,obtained from cadence, given a crank length\,---\,and the force applied to pedals.
%%%%%%%%%%%%%%%%%%%%%%%%%
\begin{figure}[h]
\centering
\includegraphics[scale=0.35]{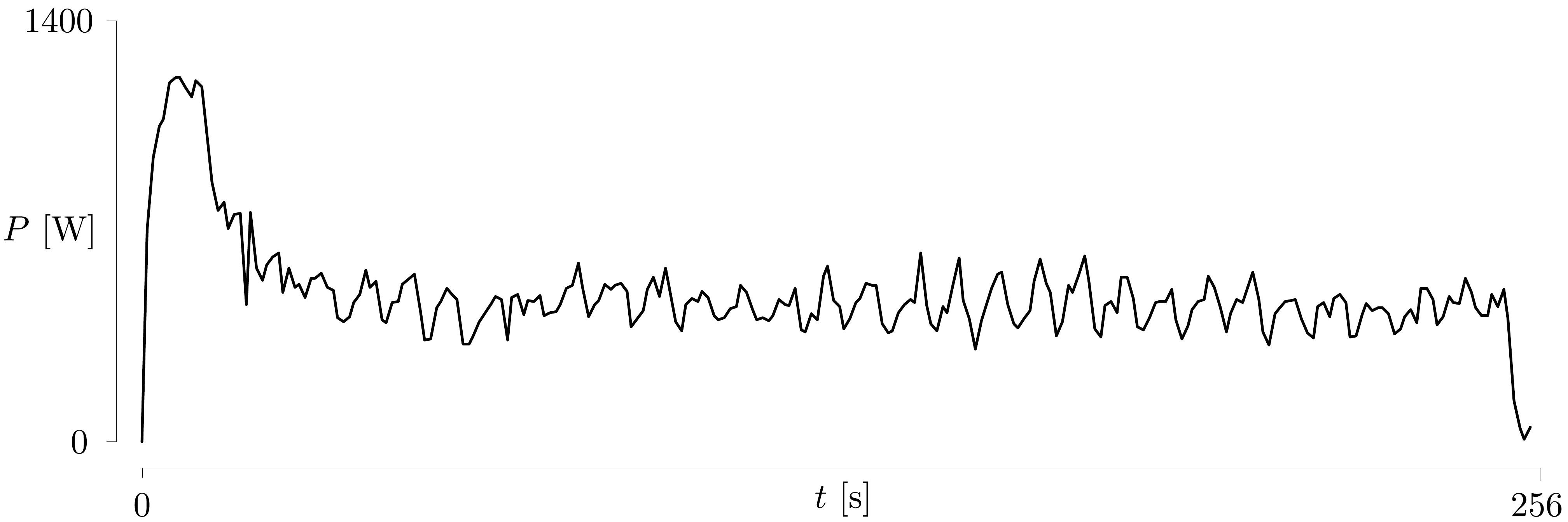}
\caption{\small Measured power,~$P$\,, as a function of the pursuit time,~$t$}
\label{fig:FigIPPower}
\end{figure}
%%%%%%%%%%%%%%%%%%%%%%%%%

\begin{figure}[h]
\centering
\includegraphics[scale=0.35]{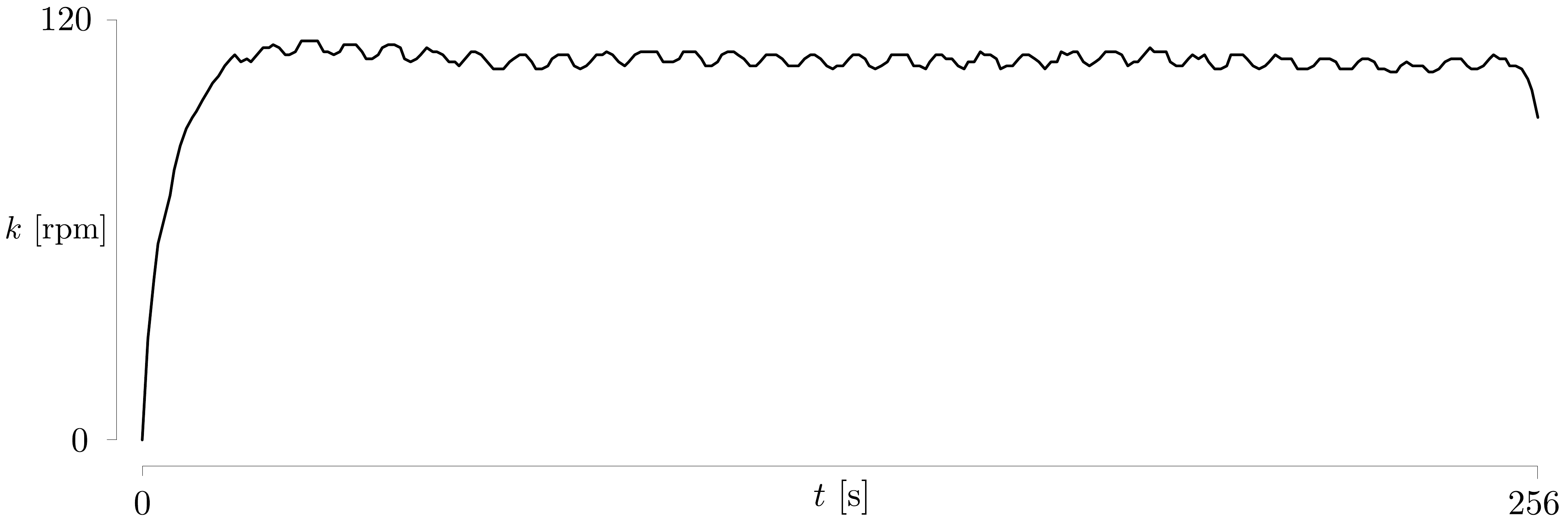}
\caption{\small Measured cadence,~$k$\,, in revolutions per minute, [rpm], as a function of the pursuit time,~$t$}
\label{fig:FigIPCadence}
\end{figure}

The measurements of power, shown in Figure~\ref{fig:FigIPPower}, oscillate about a nearly constant value, except for the initial part, which corresponds to acceleration, and the final part, where the cyclist begins to decelerate.
These oscillations are due to the repetition of straights and curves along a lap.
In particular, Figure~\ref{fig:FigIPCadence}, below, exhibits a regularity corresponding to thirty-two curves along which the cadence, and\,---\,equivalently\,---\,the wheel speed, reaches a maximum.
There are also fluctuations due to measurement errors.
A comparison of Figures~\ref{fig:FigIPPower} and \ref{fig:FigIPCadence} illustrates that power is necessarily more error sensitive than cadence, since the cadence itself is used in calculations to obtain power.
This extra sensitivity is due to intrinsic difficulties of the measurement of applied force and, herein, to the fact that values are stated at one-second intervals only, which makes them correspond to different points along the pedal rotation \citep[see also][Appendix~A]{DSSbici1}.
To diminish this effect, it is common to use a moving average, with a period of several seconds, to obtain the values of power.

To use the model to relate power and cadence, let us consider a $4000$\,-metre individual pursuit.
The model parameters are $h=1.1~\rm{m}$\,, $m=85.6~\rm{kg}$\,, ${\rm C_{d}A}=0.17~\rm{m^2}$\,, ${\rm C_{rr}}=0.0017$\,, ${\rm C_{sr}}=0.0025$\,, $\lambda=0.02$\,, $g=9.81~\rm{m/s^2}$\,, $\rho=1.17~\rm{kg/m^3}$\,.
If we use, as input, $P=488.81~\rm{W}$\,---\,which is the average of values measured over the entire pursuit\,---\,the retrodiction provided by the model results in $\overline{v}=16.86~\rm{m/s}$\,.

Let us compare this retrodiction to measurements using the fact that\,---\,for a fixed-wheel drivetrain\,---\,cadence allows us to calculate the bicycle wheel speed.
The average of the measured cadence, shown in Figure~\ref{fig:FigIPCadence}, is $k=106.56~\rm rpm$\,, which\,---\,given the gear of $9.00~\rm m$\,, over the pursuit time of $256~\rm s$\,---\,results in a distance of $4092~\rm m$\,.
Hence, the average wheel speed is~$15.98~\rm{m/s}$\,.%
\footnote{The average wheel speed is distinct from the average black-line speed,~$\overline{v}=15.63~\rm{m/s}$\,.}
%%%%%%%%%%%%%%%%%%%%%%%%%
%%%%%%%%%%%%%%%%%%%%%%%%%

The average values of the retrodicted and measured speeds appear to be sufficiently close to each other to support the empirical adequacy of our model, for the case in which its assumptions, illustrated in Figure~\ref{fig:FigBlackLine}\,---\,namely, a constant aerodynamic position and  the trajectory along the black line\,---\,are, broadly speaking, satisfied.

Specifically, they are not satisfied on the first lap, during the acceleration.
Nor can we expect them to be fully satisfied along the remainder of the pursuit, as illustrated by empirical distance of $4092\,{\rm m}>4000\,{\rm m}$\,, which indicates the deviation from the black-line trajectory.
%%%%%%%%%%%%%%%%%%%%%%%%%
\begin{figure}[h]
\centering
\includegraphics[scale=0.35]{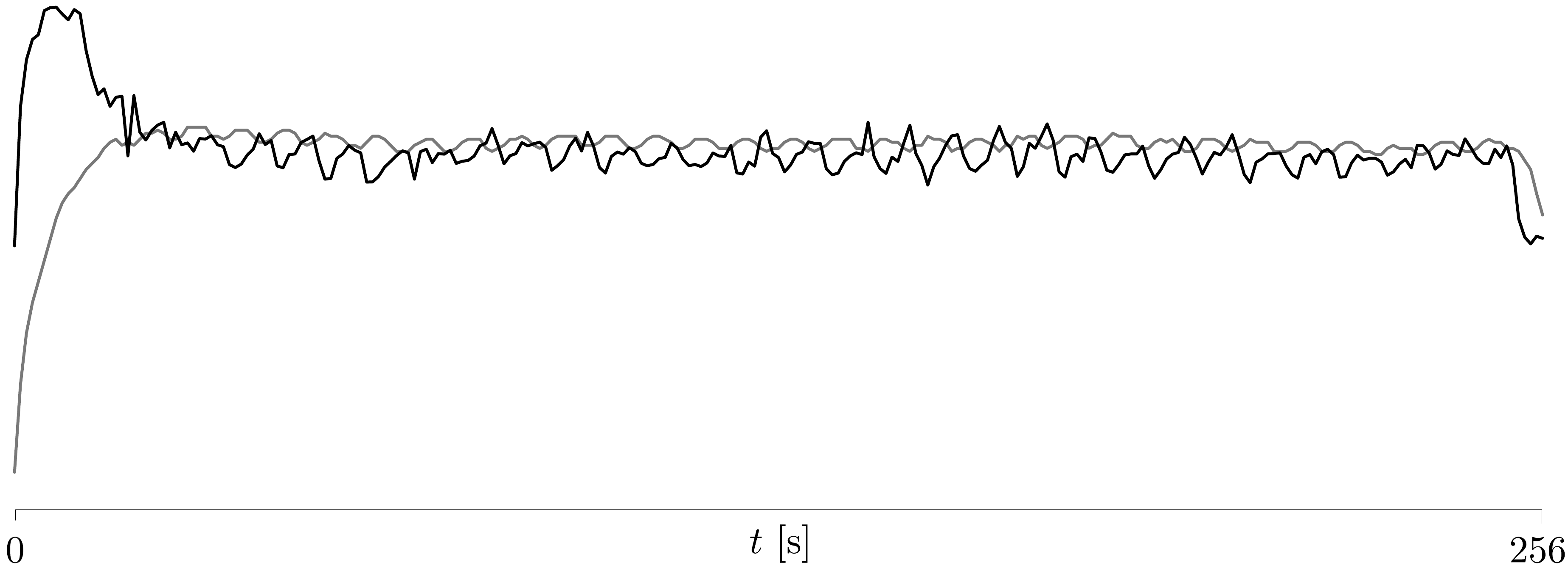}
\caption{\small Scaled values of power (black) and cadence (grey) as functions of the pursuit time,~$t$}
\label{fig:FigIPShift}
\end{figure}
%%%%%%%%%%%%%%%%%%%%%%%%%

\begin{figure}[h]
\centering
\includegraphics[scale=0.5]{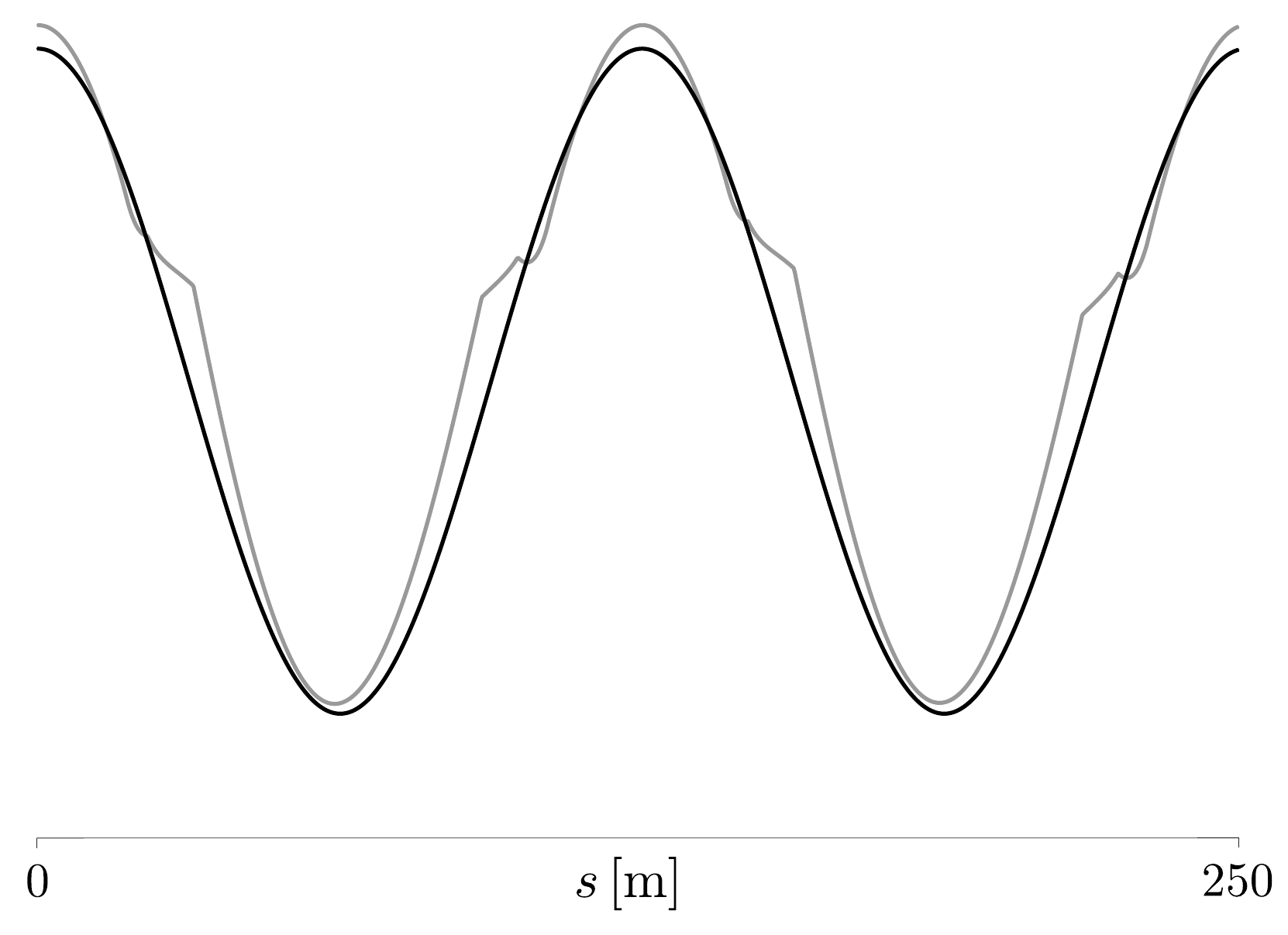}
\caption{\small Scaled values of power (black) and black-line speed (grey) as functions of the black-line distance,~$s$}
\label{fig:FigModelShift}
\end{figure}

Furthermore, Figure~\ref{fig:FigIPShift}, which is a superposition of scaled values from Figures~\ref{fig:FigIPPower} and \ref{fig:FigIPCadence}, shows the oscillations of power and cadence to be half a cycle out of phase.
However, in Figure~\ref{fig:FigModelShift} these quantities are in phase.
Therein, as input, we use simulated values of power along a lap\,---\,in a manner consistent with the measured power\,---\,as opposed to single value of an average.
Thus, according to the model, the power and the black-line speed\,---\,whose pattern within the model for a fixed-wheel drivetrain is the same as for cadence\,---\,do not exhibit the phase shift seen in the data.
%%%%%%%%%%%%%%%%%%%%%%%%%
%%%%%%%%%%%%%%%%%%%%%%%%%

The phase shift observed in Figure~\ref{fig:FigIPShift} is a result of the difference between the wheel speed, which corresponds to cadence, and the the centre-of-mass speed, which has the dominant effect on the required power.
In other words, as indicated by regularity corresponding to the thirty-two pairs, it is a result of the pattern of straights and curves on the velodrome.
It suggests that the model being considered is adequate for modelling average power, since the values of power and cadence in Figures~\ref{fig:FigIPPower} and \ref{fig:FigIPCadence} oscillate about means that are nearly constant, but less so for describing instantaneous effects.

Also, measurements of the effect caused by the difference between the wheel and the centre-of-mass speeds are affected by a fixed-wheel drivetrain.
The value of power, at each instant, is obtained from the product of measurements of~$f_{\circlearrowright}$\,, which is the force applied to pedals, and~$v_{\circlearrowright}$\,, which is the circumferential speed of the pedals \citep[e.g.,][expression~(1)]{DSSbici1},
\begin{equation}
\label{eq:PowerMeter}
P=f_{\circlearrowright}\,v_{\circlearrowright}\,.
\end{equation}
For a fixed-wheel drivetrain, there is a one-to-one relation between $v_{\circlearrowright}$ and the wheel speed.
Hence\,---\,in contrast to a free-wheel drivetrain, for which $f_{\circlearrowright}\to0\implies v_{\circlearrowright}\to0$\,---\,the momentum of a launched bicycle-cyclist system might contribute to the value of~$v_{\circlearrowright}$\,, which is tantamount to contributing to the value of cadence.

Nevertheless, even with the above caveats, the agreement between the average values of the retrodiction and measurements appears to be satisfactory.
Notably, excluding the first and last laps would increase this agreement.
For instance, if we consider, say, $33\,{\rm s} < t < 233\,{\rm s}$\,---\,with the start and end locations not at the same point, which has a negligible effect over many laps\,---\,the average power and cadence are $455.02\,{\rm W}$ and $108.78\,{\rm rpm}$\,, respectively.
Hence, the retrodicted and measured speeds are $16.45\,\rm{m/s}$ and $16.32\,\rm{m/s}$\,, respectively.
%%%%%%%%%%%%%%%%%%%%%%%%%
\begin{figure}[h]
\centering
\includegraphics[scale=0.35]{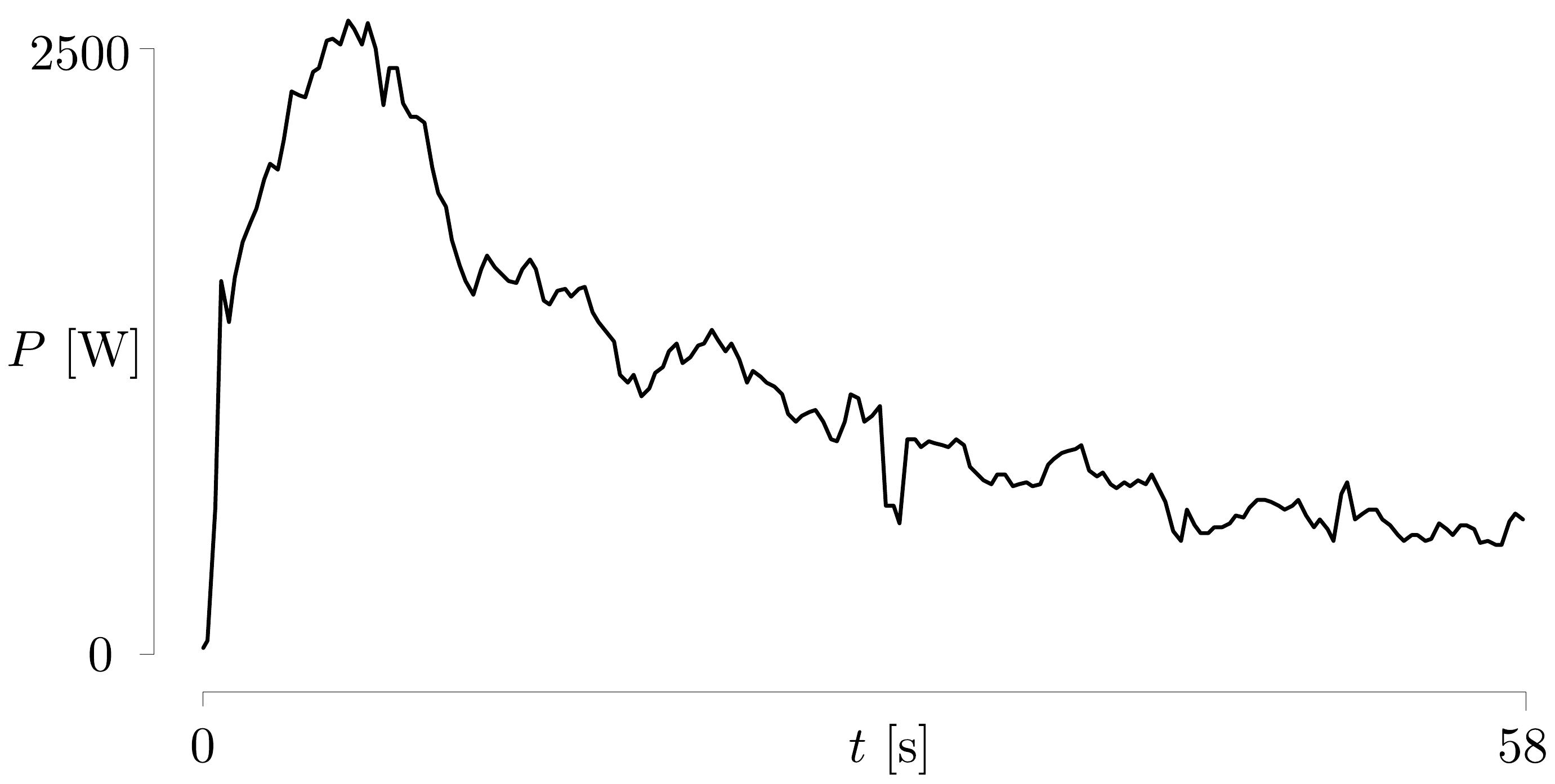}
\caption{\small Measured power,~$P$\,, as a function of the `kilo' time,~$t$}
\label{fig:FigKiloPower}
\end{figure}
%%%%%%%%%%%%%%%%%%%%%%%%%
%%%%%%%%%%%%%%%%%%%%%%%%%
\begin{figure}[h]
\centering
\includegraphics[scale=0.35]{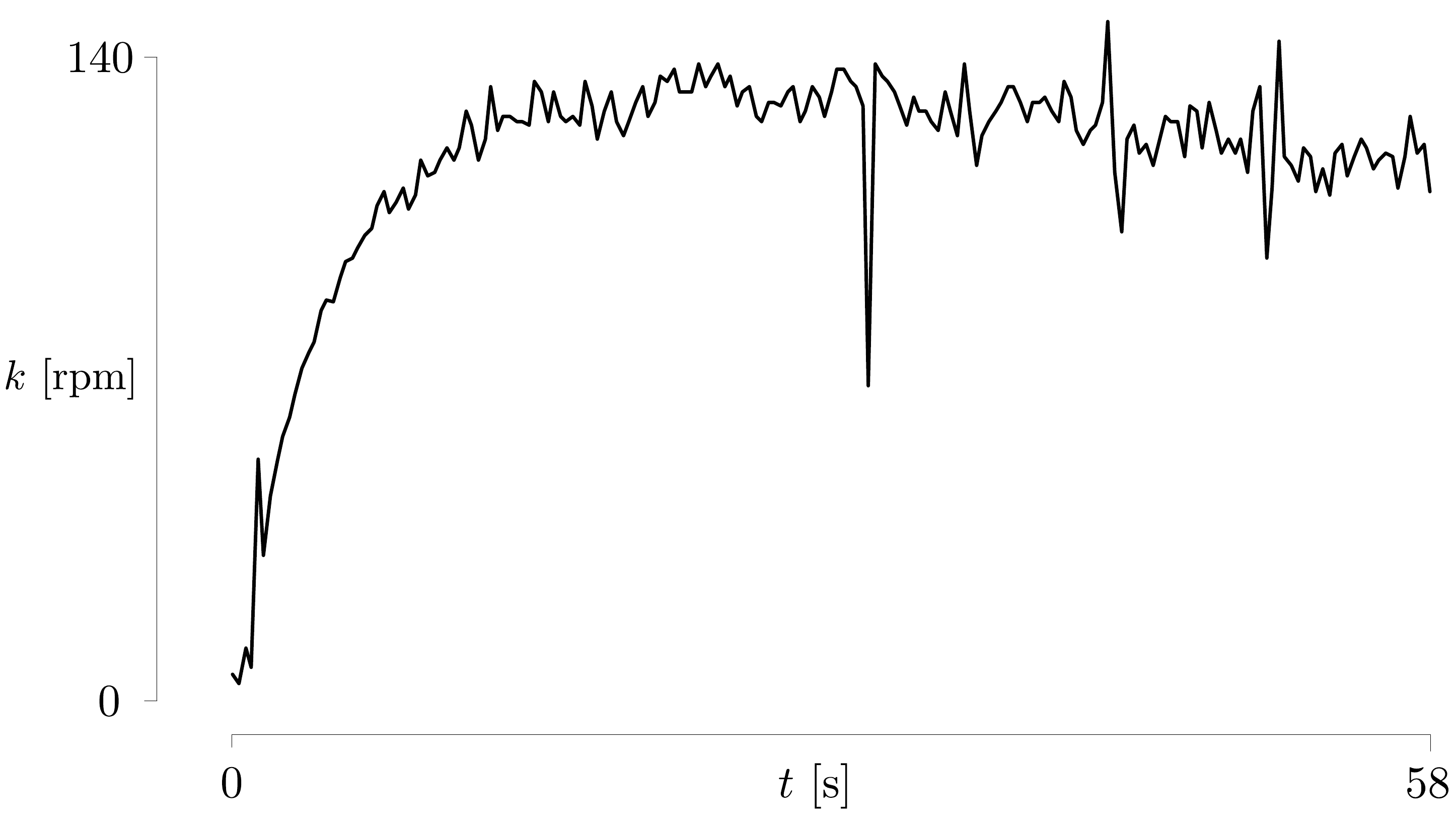}
\caption{\small Measured cadence,~$k$\,, in revolutions per minute, [rpm], as a function of the `kilo' time,~$t$}
\label{fig:FigKiloCadence}
\end{figure}
%%%%%%%%%%%%%%%%%%%%%%%%%

To illustrate limitations of the model, Figures~\ref{fig:FigKiloPower} and \ref{fig:FigKiloCadence} represent measurements for which it is not empirically adequate.
As shown in these figures, in this $1000$\,-metre time trial, commonly referred to as a `kilo', the cyclist reaches a steady cadence\,---\,and speed\,---\,with an initial output of power, in a manner similar to the one shown in Figure~\ref{fig:FigIPPower}.
Subsequently, in a manner similar to the one shown in Figure~\ref{fig:FigIPCadence}, the cadence remains almost unchanged, for the remainder of the time trial.
However, in contrast to Figure~\ref{fig:FigIPPower}, the power decreases.
Herein, as discussed by \citet[Appendix~B.2]{DSSbici2}, the cadence, as a function of time, is a consequence of both the power generated by a cyclist\,---\,at each instant\,---\,and the momentum of the moving bicycle-cyclist system, gained during the initial acceleration, which propels the pedals.
In other words, the cadence at a given instant is not solely due to the power at that instant but also depends on power expended earlier.
This situation is in contrast to the case of a steady effort in a $4000$\,-metre individual pursuit.

Given a one-to-one relationship between cadence and power, Figures~\ref{fig:FigIPPower} and \ref{fig:FigIPCadence} would remain similar for a free-wheel drivetrain, provided a cyclists keeps on pedalling in a continuous and steady manner.
Figures~\ref{fig:FigKiloPower} and \ref{fig:FigKiloCadence} would not.
In particular, Figure~\ref{fig:FigKiloCadence} would show a decrease of cadence with time, even though the bicycle speed might not decrease significantly.
%%%%%%%%%%%%%%%%%%%%%%%%%
\section{Discussion and conclusions}
\label{sec:DisCon}
%%%%%%%%%%%%%%%%%%%%%%%%%
The mathematical model presented in this article offers the basis for a quantitative study of individual time trials on a velodrome.
The model can be used to predict or retrodict the laptimes, from the measurements of power, or to estimate the power from the recorded times.
Comparisons of such predictions or retrodictions with the measurements of time, speed, cadence and power along the track offer an insight into the empirical adequacy of a model.
Given a satisfactory adequacy and appropriate measurements, the model lends itself to estimating  the rolling-resistance, lateral-friction, air-resistance and drivetrain-resistance coefficients.
One can examine the effects of power on speed and {\it vice versa}, as well as of other parameters, say, the effects of air resistance on speed.
One can also estimate the power needed for a given rider to achieve a particular result.

Furthermore, presented results allow us to comment on aspects of the velodrome design.
As illustrated in Figures~\ref{fig:FigLeanAngle}--\ref{fig:FigPower}, \ref{fig:FigCoMSpeed2}, \ref{fig:FigBLspeed}, \ref{fig:FigModelShift}, the transitions\,---\,between the straights and the circular arcs\,---\,do not result in smooth functions for the lean angles, speeds and powers.
It might suggest that a commonly used Euler spiral, illustrated in Figure~\ref{fig:FigCurvature}, is not the optimal transition curve.
Perhaps, the choice of a transition curve should consider such phenomena as the jolt, which is the temporal rate of change of acceleration.
It might also suggest the necessity for the lengthening of the transition curve.

An optimal velodrome design would strive to minimize the separation between the zero line and the curve in Figure~\ref{fig:FigAngleDiff}, which is tantamount to optimizing the track inclination to accommodate the lean angle of a rider.
The smaller the separation, the smaller the second summand in term~(\ref{eq:modelB}).
As the separation tends to zero, so does the summand.

These considerations are to be examined in future work.
Also, the inclusion, within the model, of a change of mechanical energy, discussed by \citet[Appendix~A]{BSSSbici4}, is an issue to be addressed.
Another consideration to be examined is the discrepancy between the model and measurements with respect to the phase shift between power and cadence, illustrated in Figures~\ref{fig:FigIPShift} and \ref{fig:FigModelShift}.
A possible venue for such a study is introduced by \citet[Appendix~B]{BSSSbici4}.

In conclusion, let us emphasize that our model is phenomenological.
It is consistent with\,---\,but not derived from\,---\,fundamental concepts.
Its purpose is to provide quantitative relations between the model parameters and observables.
Its key justification is the agreement between measurements and predictions or retrodictions, as discussed in Section~\ref{sec:Adequacy}.
%%%%%%%%%%%%%%%%%%%%%%%%%
\section*{Acknowledgements}
%%%%%%%%%%%%%%%%%%%%%%%%%
We wish to acknowledge Mehdi Kordi, for information on the track geometry and for measurements, used in Sections~\ref{sec:Formulation} and \ref{sec:Adequacy}, respectively;
Tomasz Danek, for statistical insights into these measurements;
Elena Patarini, for her graphic support;
Roberto Lauciello, for his artistic contribution;
Favero Electronics for inspiring this study by their technological advances of power meters.
%%%%%%%%%%%%%%%%%%%%%%%%%
\bibliographystyle{apa}
\bibliography{BSSSvelo.bib}
%%%%%%%%%%%%%%%%%%%%%%%%%
\end{document}